\documentclass[twocolumn,english,aps,prb,twocolumn,groupedadress,superscriptaddress,amssymb,amsmath]{revtex4}
\usepackage[T1]{fontenc}
\usepackage[utf8]{inputenc}
\usepackage{amsmath}
\usepackage{amssymb}
\usepackage{graphicx}
\usepackage{esint}

\makeatletter
\@ifundefined{textcolor}{}
{%
 \definecolor{BLACK}{gray}{0}
 \definecolor{WHITE}{gray}{1}
 \definecolor{RED}{rgb}{1,0,0}
 \definecolor{GREEN}{rgb}{0,1,0}
 \definecolor{BLUE}{rgb}{0,0,1}
 \definecolor{CYAN}{cmyk}{1,0,0,0}
 \definecolor{MAGENTA}{cmyk}{0,1,0,0}
 \definecolor{YELLOW}{cmyk}{0,0,1,0}
 }

\usepackage{babel}

\makeatother

\usepackage{babel}
\begin{document}

\title{Spin-thermopower in interacting quantum dots}

\author{Tomaž Rejec}

\affiliation{Faculty for Mathematics and Physics, University of Ljubljana, Jadranska~19,
Ljubljana, Slovenia}

\affiliation{Jožef Stefan Institute, Jamova~39, Ljubljana, Slovenia}

\author{Rok Žitko}

\affiliation{Jožef Stefan Institute, Jamova~39, Ljubljana, Slovenia}

\affiliation{Faculty for Mathematics and Physics, University of Ljubljana, Jadranska~19,
Ljubljana, Slovenia}

\author{Jernej Mravlje}

\affiliation{Collège de France, 11 place Marcelin Berthelot, 75005 Paris, France}

\affiliation{Centre de Physique Théorique, École Polytechnique, CNRS, 91128 Palaiseau
Cedex, France}

\affiliation{Jožef Stefan Institute, Jamova~39, Ljubljana, Slovenia}

\author{Anton Ramšak}

\affiliation{Faculty for Mathematics and Physics, University of Ljubljana, Jadranska~19,
Ljubljana, Slovenia}

\affiliation{Jožef Stefan Institute, Jamova~39, Ljubljana, Slovenia}
\begin{abstract}
Using analytical arguments and the numerical renormalization group
method we investigate the spin-thermopower of a quantum dot in a magnetic
field. In the particle-hole symmetric situation the temperature difference
applied across the dot drives a pure spin current without accompanying
charge current. For temperatures and fields at or above the Kondo
temperature, but of the same order of magnitude, the spin-Seebeck
coefficient is large, of the order of $k_{B}/|e|$. Via a mapping,
we relate the spin-Seebeck coefficient to the charge-Seebeck coefficient
of a negative-$U$ quantum dot where the corresponding result was
recently reported by Andergassen \textsl{et al.} in Phys. Rev. B \textbf{84},
241107 (2011). For several regimes we provide simplified analytical
expressions. In the Kondo regime, the dependence of the spin-Seebeck
coefficient on the temperature and the magnetic field is explained
in terms of the shift of the Kondo resonance due to the field and
its broadening with the temperature and the field. We also consider
the influence of breaking the particle-hole symmetry and show that
a pure spin current can still be realized provided a suitable electric
voltage is applied across the dot. Then, except for large asymmetries,
the behavior of the spin-Seebeck coefficient remains similar to that
found in the particle-hole symmetric point.
\end{abstract}
\maketitle

\section{Introduction}

Thermoelectricity is the occurrence of electric voltages in the presence
of temperature differences or vice-versa. Devices based on thermoelectric
phenomena can be used for several applications including power generation,
refrigeration, and temperature measurement.\cite{Mahan97} Thermoelectric
phenomena are also of fundamental scientific interest as they reveal
information about a system which is unavailable in standard charge
transport measurements.\cite{Segal05} Recent reinvigoration in this
field is due to a large thermoelectric response found in some strongly
correlated materials, \textit{e.g.} sodium cobaltate,\cite{Terasaki97}
as well as due to the investigation of thermopower in nanoscale junctions
such as quantum point contacts, silicon nanowires, carbon nanotubes,
and molecular junctions.\cite{Dubi11,Rejec02} Recently, thermopower
of Kondo correlated quantum dots has been measured\cite{Scheibner05}
and theoretically analyzed.\cite{Costi10a,Andergassen11}

Importantly, the thermoelectric effects turned to be useful\cite{Uchida08}
also for spintronics\cite{Zutic04}. Spintronic devices exploiting
the spin degree of freedom of an electron, such as the prototypical
Datta-Das spin field-effect transistor,\cite{Datta90} promise many
advantages over the conventional charge-based electronic devices,
most notably lower power consumption and heating.\cite{Wolf01} Recently,
the spin-Hall effect was utilized to realize the spin transistor idea
experimentally.\cite{Wunderlich10} Generating spin currents plays
a fundamental role in driving spintronic devices. Several methods
of generating spin currents have been put forward such as electrical
based on the tunneling from ferromagnetic contacts and optical based
on excitation of carriers in semiconductors with circularly polarized
light.\cite{Zutic04} Another possibility is thermoelectrical injection.
In a recent breakthrough, the spin-Seebeck effect, where the spin
current is driven by a temperature difference across the sample, has
been observed in a metallic ferromagnet and suggested as a spin current
source.\cite{Uchida08}

\begin{figure}
\centering{}\includegraphics[width=0.85\columnwidth]{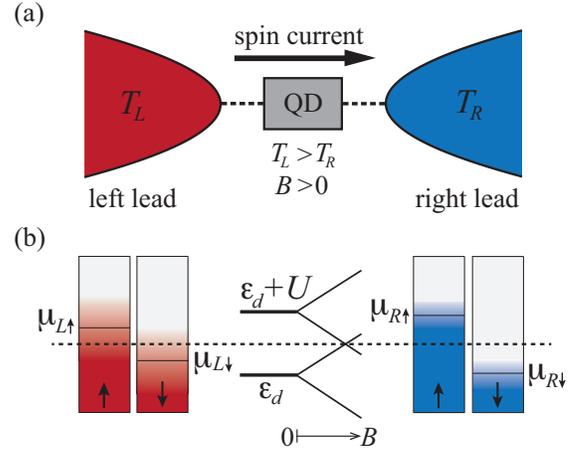} \caption{\label{schem}(a) Quantum dot attached to paramagnetic leads held
at different temperatures. (b) Energy levels in the leads and in the
quantum dot (at the particle-hole symmetric point, as a function of
the magnetic field). }
\end{figure}

Some of the above mentioned methods have an undesirable property that
the generated spin current is accompanied by an electrical current,
which leads to dissipation and heat. In this paper we consider a device,
which uses the spin-Seebeck effect to generate a pure spin current.
It consists of a quantum dot in a magnetic field attached to paramagnetic
leads, as shown in Fig.~\ref{schem}(a). The leads are held at different
temperatures. In the particle-hole symmetric situation, this setup
generates a pure spin current across the quantum dot. Away from the
particle-hole symmetric point, a pure spin current can be generated
provided a suitable electrical voltage is applied. By tuning the temperature
and the magnetic field, large values of the spin-Seebeck coefficient
of the order of $k_{B}/|e|$ can be reached. For the particle-hole
symmetric case this was first recognized in a related problem of the
charge thermopower in a negative-$U$ quantum dot by Andergassen \textsl{et
al.} in Ref.~{[}\onlinecite{Andergassen11}{]}.

The outline of the paper is as follows. First, in Secs.~\ref{sec:Model}
and \ref{sec:Spin-Seebeck-coefficient}, we present the model and
introduce the notion of spin-thermopower and define the spin-Seebeck
coefficient. In Sec.~\ref{sec:Noninteracting-quantum-dot}, we present
exact results for a device containing a noninteracting quantum dot.
Then, in Sec.~\ref{sec:Method}, we describe the details of the numerical
renormalization group (NRG) calculation and, in Sec.~\ref{sec:Results},
we present the NRG results for a device with an interacting quantum
dot. We also derive analytical expressions for the spin-Seebeck coefficient
in several asymptotic regimes and find the boundaries between those
regimes. Finally, in Sec.~\ref{sec:Summary}, we give conclusions
and examine whether our results can be easily observed experimentally.
In Appendix A we present an intuitive explanation why a pure spin
current is generated in the particle-hole symmetric point.

The spin-Seebeck coefficient was calculated for a single-molecule-magnet
attached to metallic electrodes\cite{Wang10} and for a quantum dot
in contact with ferromagnetic electrodes,\cite{Dubi09} both within
the sequential tunneling approximation. The later system was also
analyzed using the equation of motion technique\cite{Swirkowicz09}
as well as the Hartree-Fock approximation.\cite{Ying10} All these
methods fail to describe correctly the effects of Kondo correlations
occurring at low temperatures and magnetic fields in such systems.
Thermospin effects in quasi-one dimensional quantum wires in the presence
of a magnetic field and Rashba spin-orbit interaction were also studied
theoretically.\cite{Lu10}

\section{\label{sec:Model}Model}

Our starting point is the standard Anderson impurity Hamiltonian\cite{Hewson93}

\begin{eqnarray}
H & = & \sum_{\sigma}\left(\epsilon_{d}+\sigma\frac{B}{2}\right)n_{d\sigma}+Un_{d\uparrow}n_{d\downarrow}+\nonumber \\
 &  & +\sum_{k\alpha\sigma}\varepsilon_{k\alpha}c_{k\alpha\sigma}^{\dagger}c_{k\alpha\sigma}+\sum_{k\alpha\sigma}V_{k\alpha}c_{k\alpha\sigma}^{\dagger}d_{\sigma}+\mathrm{h.c.}\label{eq:ham}
\end{eqnarray}
Here we assume that only the highest occupied energy level $\epsilon_{d}$
in the dot is relevant for the transport properties. Due to the electron's
spin $\sigma=\pm1$, in external magnetic field this level is split
by the Zeeman energy $g\mu_{B}B$. Rather than by an external magnetic
field, the dot level could also be split by a local exchange field
due to an attached ferromagnetic electrode.\cite{Martinek05} To keep
the notation short, in Eq.~(\ref{eq:ham}) we express the magnetic
field in energy units, \textit{i.e.}, $g\mu_{B}=1$. $U$ is the Coulomb
charging energy of the dot, while $\Gamma=\sum_{\alpha}\Gamma_{\alpha}$
is the total hybridization, where $-i\Gamma_{\alpha}=\sum_{k}\left|V_{k\alpha}\right|^{2}/\left(\omega-\epsilon_{k\alpha}+i\delta\right)$
is the hybridization function of the dot level with the states in
the left ($\alpha=L$) and the right ($\alpha=R$) lead, which we
assume to be energy independent, as appropriate for wide bands in
leads with constant densities of states. Furthermore, we will mostly
consider the system at the particle-hole symmetric point (in experiments,
the deviation from particle-hole symmetry is easily controlled by
the gate voltage) where the dot level is at $\epsilon_{d}=\mu-\frac{U}{2}$
in equilibrium. In what follows we set the equilibrium chemical potential
$\mu$ to zero. In such a regime the spin of the quantum dot is quenched
at temperatures (again we use the energy units, $k_{B}=1$) and magnetic
fields low compared to the Kondo temperature\cite{Horvatic85} 
\begin{equation}
T_{K}=\sqrt{\frac{U\Gamma}{2}}e^{-\frac{\pi U}{8\Gamma}+\frac{\pi\Gamma}{2U}}.\label{eq:Tk}
\end{equation}
To describe the system away from the particle-hole symmetric point
we introduce the asymmetry parameter $\delta=\epsilon_{d}+\frac{U}{2}$.

\section{\label{sec:Spin-Seebeck-coefficient}Spin-Seebeck coefficient}

In a non-equilibrium situation the distribution of electrons with
spin $\sigma$ in the lead $\alpha$ is described by the Fermi-Dirac
distribution function $f_{\alpha\sigma}\left(\omega\right)=1/\left[e^{\left(\omega-\mu_{\alpha\sigma}\right)/T_{\alpha}}+1\right]$
with $\mu_{\alpha\sigma}$ and $T_{\alpha}$ being the spin dependent
chemical potentials and the temperature in the lead $\alpha$, respectively
{[}Fig.~\ref{schem}(b){]}. The electrical current of electrons with
spin $\sigma$ is 
\begin{equation}
I_{\sigma}=\frac{e}{h}\int\mathrm{d}\omega\left[f_{L\sigma}\left(\omega\right)-f_{R\sigma}\left(\omega\right)\right]{\cal T}_{\sigma}\left(\omega\right),\label{eq:I_sigma}
\end{equation}
 where $e$ is the electron charge, $h$ is the Planck constant, and
${\cal T}_{\sigma}\left(\omega\right)=\pi\frac{2\Gamma_{L}\Gamma_{R}}{\Gamma_{L}+\Gamma_{R}}A_{\sigma}\left(\omega\right)$
is the transmission function of electrons with spin $\sigma$.\cite{Meir92}
It is calculated from the impurity spectral function $A_{\sigma}\left(\omega\right)=-\frac{1}{\pi}\mathrm{Im}G_{\sigma}\left(\omega\right)$
where 
\[
G_{\sigma}\left(\omega\right)=\frac{1}{\omega-\epsilon_{d}-\sigma\frac{B}{2}+i\Gamma-\Sigma_{\sigma}\left(\omega\right)}
\]
 is the retarded impurity Green's function and the interaction self-energy
$\Sigma_{\sigma}\left(\omega\right)$ accounts for the many-body effects.
Introducing the average chemical potential in each of the leads, 
\[
\mu_{\alpha}=\frac{1}{2}\left(\mu_{\alpha\uparrow}+\mu_{\alpha\downarrow}\right),
\]
 we can parametrize the temperatures and the chemical potentials in
the leads in terms of their average values 
\[
T=\frac{1}{2}\left(T_{L}+T_{R}\right)
\]
 and 
\[
\mu=\frac{1}{2}\left(\mu_{L}+\mu_{R}\right)=0,
\]
 the temperature difference 
\[
\Delta T=T_{L}-T_{R},
\]
 the voltage 
\[
eV=\mu_{L}-\mu_{R},
\]
 and the spin voltage 
\[
eV_{s}=\left(\mu_{L\uparrow-}\mu_{L\downarrow}\right)-\left(\mu_{R\uparrow-}\mu_{R\downarrow}\right).
\]
 Assuming $\Delta T$, $eV$ and $eV_{s}$ to be small, the electrical
current $I=I_{\uparrow}+I_{\downarrow}$ 
\begin{equation}
\begin{alignedat}{1}I=\frac{e}{h} & \left[\left({\cal I}_{1\uparrow}+{\cal I}_{1\downarrow}\right)\frac{\Delta T}{T}+\left({\cal I}_{0\uparrow}+{\cal I}_{0\downarrow}\right)eV+\phantom{\frac{1}{2}}\right.\\
 & \left.+\frac{1}{2}\left({\cal I}_{0\uparrow}-{\cal I}_{0\downarrow}\right)eV_{s}\right]
\end{alignedat}
\label{eq:I}
\end{equation}
 and the spin current $I_{s}=I_{\uparrow}-I_{\downarrow}$ 
\begin{equation}
\begin{alignedat}{1}I_{s}=\frac{e}{h} & \left[\left({\cal I}_{1\uparrow}-{\cal I}_{1\downarrow}\right)\frac{\Delta T}{T}+\left({\cal I}_{0\uparrow}-{\cal I}_{0\downarrow}\right)eV+\phantom{\frac{1}{2}}\right.\\
 & \left.+\frac{1}{2}\left({\cal I}_{0\uparrow}+{\cal I}_{0\downarrow}\right)eV_{s}\right]
\end{alignedat}
\label{eq:Is}
\end{equation}
 can both be expressed in terms of the transport integrals 
\begin{equation}
{\cal I}_{n\sigma}=\int\mathrm{d}\omega\omega^{n}\left[-f^{\prime}\left(\omega\right)\right]{\cal T}_{\sigma}\left(\omega\right)\label{eq:transint}
\end{equation}
 where $f\left(\omega\right)$ is the Fermi-Dirac function at $T$
and $\mu$.

At the particle-hole symmetric point one has ${\cal I}_{0\uparrow}={\cal I}_{0\downarrow}\equiv{\cal I}_{0}$
and ${\cal I}_{1\uparrow}=-{\cal I}_{1\downarrow}\equiv{\cal I}_{1}$
due to the symmetry of the spectral functions, $A_{\downarrow}\left(\omega\right)=A_{\uparrow}\left(-\omega\right)$.
Thus, 
\begin{eqnarray*}
I & = & \frac{2e}{h}{\cal I}_{0}eV,\\
I_{s} & = & \frac{2e}{h}\left({\cal I}_{1}\frac{\Delta T}{T}+\frac{1}{2}{\cal I}_{0}eV_{s}\right).
\end{eqnarray*}
In the absence of a voltage applied across the dot only the spin current
will flow. In Appendix A we present an intuitive explanation of this
result. 

The spin current is thus driven by the temperature difference and
the spin voltage. The appropriate measure for the spin thermopower,
\textsl{i.e.}, the ability of a device to convert temperature difference
to spin voltage, is the spin-Seebeck coefficient $S_{s}$, expressed
in what follows in units of $k_{B}/\left|e\right|$, given by the
ratio of the two driving forces required for the spin current to vanish,
\begin{equation}
S_{s}=-\left.\frac{eV_{s}}{\Delta T}\right|_{I_{s}=0}=\frac{2}{T}\frac{{\cal I}_{1}}{{\cal I}_{0}}.\label{eq:S}
\end{equation}
Note that an asymmetry in the coupling to the leads, $\Gamma_{L}\ne\Gamma_{R}$,
does not influence the value of the spin-Seebeck coefficient provided
$\Gamma=\Gamma_{L}+\Gamma_{R}$ stays constant.

Through a mapping that interchanges the spin and pair degrees of freedom,
$d_{\downarrow}^{\dagger}\rightarrow d_{\downarrow}$ and $c_{k\downarrow}^{\dagger}\rightarrow c_{-k\downarrow}$,
$ $the Hamiltonian (\ref{eq:ham}) at the particle-hole symmetric
point in a magnetic field $B$ is isomorphic to a negative-$U$ Hamiltonian
away from the particle-hole symmetric point and in the absence of
the magnetic field.\cite{Taraphder91,Mravlje05,Koch07} The spin-down
spectral function transforms as $A_{\downarrow}\left(\omega\right)\rightarrow A_{\downarrow}\left(-\omega\right)=A_{\uparrow}\left(\omega\right)$.
Consequently, the charge thermopower in the transformed model, 
\[
S=-\left.\frac{eV}{\Delta T}\right|_{I=0}=\frac{1}{T}\frac{{\cal I}_{1}}{{\cal I}_{0}},
\]
is (up to a factor of two, which could be absorbed in a redefinition
of the spin-voltage) the same as the spin thermopower in the original
model. If transformed appropriately the results for the particle-hole
symmetric case presented in this paper agree with those reported earlier
for the charge thermopower in a negative-$U$ quantum dot.\cite{Andergassen11}

Away from the particle-hole symmetric point one needs to apply an
electrical voltage across the dot in order to make the electrical
current vanish. In this regime, the spin-Seebeck coefficient 
\begin{equation}
S_{s}=-\left.\frac{eV_{s}}{\Delta T}\right|_{I=0,I_{s}=0}=\frac{1}{T}\left(\frac{{\cal I}_{1\uparrow}}{{\cal I}_{0\uparrow}}-\frac{{\cal I}_{1\downarrow}}{{\cal I}_{0\downarrow}}\right)\label{eq:Ss_away}
\end{equation}
and the required electrical voltage
\begin{equation}
eV=-\frac{\Delta T}{T}\left(\frac{{\cal I}_{1\uparrow}}{{\cal I}_{0\uparrow}}+\frac{{\cal I}_{1\downarrow}}{{\cal I}_{0\downarrow}}\right)\label{eq:ev_away}
\end{equation}
can be readily derived from Eqs.~(\ref{eq:I}) and (\ref{eq:Is}).
By means of the above transformation, these results can also be applied
to the case of a negative-$U$ quantum dot in a finite magnetic field.
Notice that the applied electrical voltage (charge-Seebeck coefficient)
maps to a spin-voltage (spin-Seebeck coefficient) of the negative-$U$
device.

\section{\label{sec:Noninteracting-quantum-dot}Noninteracting quantum dot}

The spin-Seebeck coefficient of a noninteracting quantum dot, $U=0$,
can be related to the Seebeck coefficient of a spinless problem with
the impurity Green's function $G\left(\omega\right)=1/\bigl(\omega-\epsilon_{d}+i\Gamma\bigr)$.
The corresponding spectral function is of a Lorentzian form. The Seebeck
coefficient can be expressed in terms of the trigamma function\cite{Abramowitz64}
$\psi^{\prime}\left(z\right)=\sum_{n=0}^{\infty}\left(z+n\right)^{-2}$,

\begin{equation}
{\cal S}\left(\epsilon_{d}\right)=2\pi\frac{\mathrm{Im}\left\{ \frac{\Gamma+i\epsilon_{d}}{2\pi T}\psi^{\prime}\left(\frac{1}{2}+\frac{\Gamma+i\epsilon_{d}}{2\pi T}\right)\right\} }{\mathrm{Re}\left\{ \psi^{\prime}\left(\frac{1}{2}+\frac{\Gamma+i\epsilon_{d}}{2\pi T}\right)\right\} },\label{eq:exact_NI0}
\end{equation}
and is an odd function of $\epsilon_{d}$, ${\cal S}\left(-\epsilon_{d}\right)=-{\cal S}\left(\epsilon_{d}\right)$.

Below we first study the particle-hole symmetric case and then generalize
the results to the asymmetric problem.

\subsection*{Particle-hole symmetric point, $\boldsymbol{\delta=0}$}

\begin{figure}
\begin{centering}
\includegraphics[width=1\columnwidth]{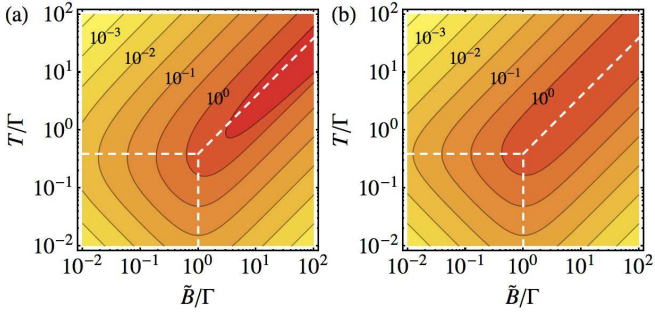} 
\par\end{centering}

\caption{\label{S0_BT}Temperature and magnetic field dependence of the spin-Seebeck
coefficient (in units of $k_{B}/\left|e\right|$) of a noninteracting
quantum dot. The maxima at $\tilde{B}=\Gamma$, $T=0.34\Gamma$ and
$T=0.34\tilde{B}$ are indicated with dashed lines. (a) Exact result,
Eq.~(\ref{eq:exact_NI}). (b) Interpolation formula, Eq.~(\ref{eq:interp_NI}).}
\end{figure}
In the absence of interaction the impurity Green's function is $G_{\sigma}\left(\omega\right)=1/\bigl(\omega-\sigma\tilde{B}+i\Gamma\bigr)$.
We introduced the energy level shift $\tilde{B}$ due to the magnetic
field. In the non-interacting case, $\tilde{B}=\frac{B}{2}$. From
Eqs.~(\ref{eq:S}) and (\ref{eq:exact_NI0}) 
\begin{equation}
S_{s}={\cal S}\bigl(\tilde{B}\bigr)-{\cal S}\bigl(-\tilde{B}\bigr)=2{\cal S}\bigl(\tilde{B}\bigr).\label{eq:exact_NI}
\end{equation}

For $T\ll\max\left(\Gamma,\tilde{B}\right)$ the spin-Seebeck coefficient
is proportional to the temperature, 
\begin{equation}
S_{s}=\frac{4\pi^{2}}{3}\frac{\tilde{B}T}{\tilde{B}^{2}+\Gamma^{2}},
\end{equation}
 a results which can also be derived by performing the Sommerfeld
expansion of the transport integrals in Eq.~(\ref{eq:transint}).
It increases linearly with the field in low magnetic fields, $\tilde{B}\ll\Gamma$,
\begin{equation}
S_{s}=\frac{4\pi^{2}}{3}\frac{\tilde{B}T}{\Gamma^{2}},\label{eq:nonint_lowBT}
\end{equation}
 and is inversely proportional to the field in high magnetic fields,
$\tilde{B}\gg\Gamma$, 
\begin{equation}
S_{s}=\frac{4\pi^{2}}{3}\frac{T}{\tilde{B}}.\label{eq:nonint_lowT}
\end{equation}
 It reaches its maximal value at $\tilde{B}=\Gamma$.

For $\tilde{B}\ll\max\left(\Gamma,T\right)$ the spin-Seebeck coefficient
is proportional to the magnetic field: 
\begin{equation}
S_{s}=\frac{2\tilde{B}}{T}\left(1+\frac{\Gamma}{2\pi T}\frac{\psi^{\prime\prime}\left(\frac{1}{2}+\frac{\Gamma}{2\pi T}\right)}{\psi^{\prime}\left(\frac{1}{2}+\frac{\Gamma}{2\pi T}\right)}\right).\label{eq:nonint_lowBx}
\end{equation}
 In the low temperature limit, $T\ll\Gamma$, where we recover Eq.~(\ref{eq:nonint_lowBT}),
it increases linearly with the temperature, reaching its maximal value
at $T=0.34\Gamma$. It is inversely proportional to the temperature
in the $T\gg\Gamma$ limit: 
\begin{equation}
S_{s}=\frac{2\tilde{B}}{T}.\label{eq:nonint_lowB}
\end{equation}

As shown in Fig.~\ref{S0_BT}, an approximation that smoothly interpolates
between the regimes of Eqs.~(\ref{eq:nonint_lowBT}), (\ref{eq:nonint_lowT})
and (\ref{eq:nonint_lowB}), 
\begin{equation}
S_{s}^{-1}=\left(\frac{4\pi^{2}}{3}\frac{\tilde{B}T}{\Gamma^{2}}\right)^{-1}+\left(\frac{4\pi^{2}}{3}\frac{T}{\tilde{B}}\right)^{-1}+\left(\frac{2\tilde{B}}{T}\right)^{-1},\label{eq:interp_NI}
\end{equation}
 is qualitatively correct also at $\tilde{B},T\sim\Gamma$. According
to this interpolation formula, the spin-Seebeck coefficient at $\tilde{B}=\Gamma$,
$T=0.34\Gamma$ where the maxima merge is $S_{s}=1.45$, which compares
well with the exact result of Eq.~(\ref{eq:exact_NI}), $S_{s}=1.61$.
From this point, the high spin-Seebeck coefficient ridge of $S_{s}\sim1$
continues to higher temperatures and fields along the $T=0.34\tilde{B}$
line. Note also that at the ridge the magnetic field and the temperature
are of the same order of magnitude.

\subsection*{Away from the particle-hole symmetric point, $\boldsymbol{\delta\ne0}$}

\begin{figure}
\begin{centering}
\includegraphics[width=1\columnwidth]{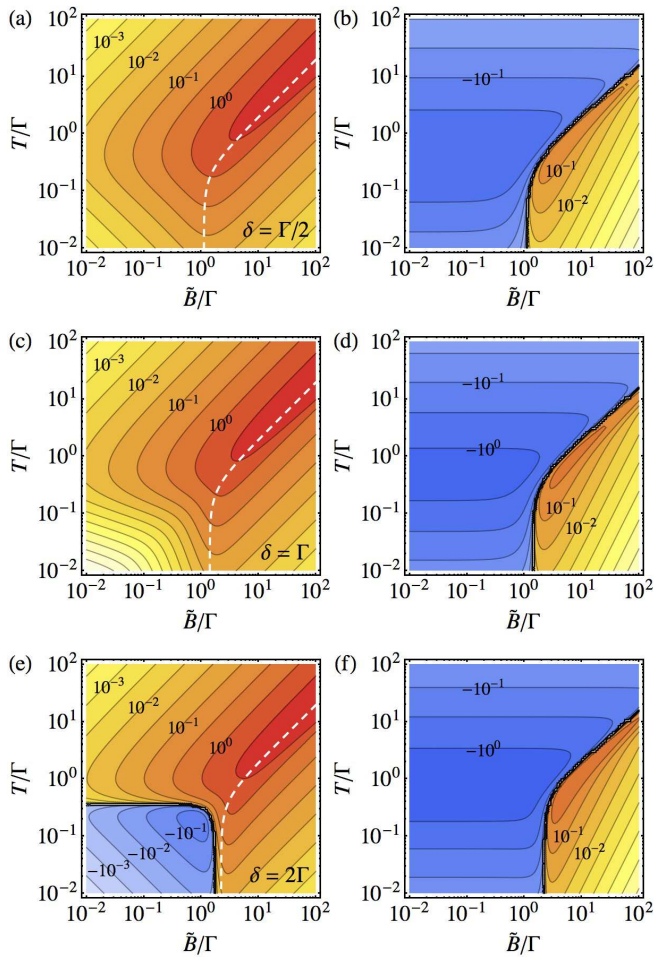} 
\par\end{centering}

\caption{\label{S0_BT_away}Temperature and magnetic field dependence of the
spin-Seebeck coefficient (in units of $k_{B}/\left|e\right|$) of
a noninteracting quantum dot away from the particle-hole symmetric
point for $ $(a) $\delta=\Gamma/2$, (c) $\delta=\Gamma$ and (e)
$\delta=2\Gamma$. The corresponding electrical voltages, $-eV/\Delta T$,
required for electrical current to vanish are shown in (b), (d) and
(f). Along the dashed lines in (a), (c) and (e) the required electrical
voltage is zero.}
\end{figure}
Here the spin-Seebeck coefficient, Eq.~(\ref{eq:Ss_away}), 
\[
S_{s}={\cal S}\bigl(\delta+\tilde{B}\bigr)-{\cal S}\bigl(\delta-\tilde{B}\bigr),
\]
and the electrical voltage required for the electrical current to
vanish, Eq.~(\ref{eq:ev_away}), 
\[
-\frac{eV}{\Delta T}={\cal S}\bigl(\tilde{B}+\delta\bigr)-{\cal S}\bigl(\tilde{B}-\delta\bigr),
\]
can also be related to the Seebeck coefficient of the spinless problem,
Eq.~(\ref{eq:exact_NI0}). 

In Figs.~\ref{S0_BT_away}(a), (c) and (e) the spin-Seebeck coefficient
$S_{s}$ is plotted for various values of the asymmetry parameter
$\delta$. For $\delta<\Gamma$ the spin-Seebeck coefficient qualitatively
resembles its $\delta=0$ behavior. At $\delta=\Gamma$ it is suppressed
in the $\Tilde B,T\ll\Gamma$ region. For $\delta>\Gamma$ the spin-Seebeck
coefficient becomes negative at low temperatures and fields. Its sign
changes at $\tilde{B},T\sim\sqrt{\delta^{2}-\Gamma^{2}}$, as one
can check by requiring ${\cal S}\bigl(\delta+\tilde{B}\bigr)={\cal S}\bigl(\delta-\tilde{B}\bigr)$
and using the approximate expression of Eq.~(\ref{eq:interp_NI})
together with Eq.~(\ref{eq:exact_NI}), ${\cal S}\left(\epsilon_{d}\right)=\frac{1}{2}\left.S_{s}\right|_{\tilde{B}\rightarrow\epsilon_{d}}$.

As evident from Figs.~\ref{S0_BT_away}(b) (d) and (f), the electrical
voltage required to stop the electrical current, $-eV/\Delta T$,
can take both positive and negative values. At a given temperature
this voltage is zero at a magnetic field $\tilde{B}$ where ${\cal S}\bigl(\tilde{B}+\delta\bigr)={\cal S}\bigl(\tilde{B}-\delta\bigr)$.
This condition is fulfilled close to the point where the spin-Seebeck
coefficient reaches a maximum as a function of the field, as shown
with dashed lines in Figs.~\ref{S0_BT_away}(a), (c) and (e).

\section{\label{sec:Method}Method}

To evaluate the transport integrals, Eq.~(\ref{eq:transint}), in
the interacting case we employed the numerical renormalization group
(NRG) method\cite{wilson1975,krishna1980a,bulla2008}. This method
allows to compute the dynamical properties of quantum impurity models
in a reliable and rather accurate way. The approach is based on the
discretization of the continuum of bath states, transformation to
a linear tight-binding chain Hamiltonian, and iterative diagonalization
of this discretized representation of the original problem.

The Lehmann representation of the impurity spectral function is\cite{bulla2008}
\[
\begin{split}A_{\sigma}\left(\omega\right) & =\frac{1}{Z}\sum_{p,r}\left(e^{-\frac{E_{p}}{T}}+e^{-\frac{E_{r}}{T}}\right)\times\\
 & \times\left|\left\langle p\left|d_{\sigma}^{\dagger}\right|r\right\rangle \right|^{2}\delta\left(\omega-\left(E_{p}-E_{r}\right)\right).
\end{split}
\]
 Here $p$ and $r$ index the many-particle levels $|p\rangle$ and
$|r\rangle$ with energies $E_{p}$ and $E_{r}$, respectively. $Z$
is the partition function 
\[
Z=\sum_{p}e^{-\frac{E_{p}}{T}}.
\]
 This can also be expressed as\cite{yoshida2009a,yoshida2009b} 
\[
\begin{split}A_{\sigma}\left(\omega\right) & =\frac{1}{Zf\left(\omega\right)}\sum_{p,r}e^{-\frac{E_{p}}{T}}\times\\
 & \times\left|\left\langle p\left|d_{\sigma}^{\dagger}\right|r\right\rangle \right|^{2}\delta\left(\omega-\left(E_{p}-E_{r}\right)\right),
\end{split}
\]
 where $f(\omega)$ is the Fermi-Dirac function. The transport integrals
(\ref{eq:transint}) are then calculated as\cite{yoshida2009a,yoshida2009b}
\begin{align}
{\cal I}_{n\sigma} & =\pi\frac{2\Gamma_{L}\Gamma_{R}}{\Gamma_{L}+\Gamma_{R}}\times\nonumber \\
 & \times\frac{1}{ZT}\sum_{p,r}\left|\left\langle p\left|d_{\sigma}^{\dagger}\right|r\right\rangle \right|^{2}\frac{\left(E_{p}-E_{r}\right)^{n}}{e^{\frac{E_{p}}{T}}+e^{\frac{E_{e}}{T}}}.\label{eq:transintnrg}
\end{align}
 In this approach it is thus not necessary to calculate the spectral
function itself, thus the difficulties with the spectral function
broadening and numerical integration do not arise. Furthermore, a
single NRG run is sufficient to obtain the transport integrals in
the full temperature interval. The calculations have been performed
using the discretization parameter $\Lambda=2$, the truncation cutoff
set at $12\omega_{N}$, and twist averaging over $N_{z}=8$ interleaved
discretization meshes. The spin U(1) and the isospin SU(2) symmetries
have been used to simplify the calculations.

An alternative approach for computing the transport integrals consists
in calculating the spectral functions using the density-matrix NRG
method\cite{hofstetter2000} or its improvements, the complete-Fock-space
NRG\cite{anders2005,anders2006,peters2006} or the full-density-matrix
(FDM) NRG.\cite{weichselbaum2007} In this case, a separate calculation
has to be performed for each temperature $T$. The approaches based
on the reduced density matrix are required to correctly determine
the \textit{high-energy parts} of the spectral function in the presence
of the magnetic field.\cite{hofstetter2000} Nevertheless, since the
main contribution to the transport integrals comes from the spectral
weight on the \textit{low-energy scale} of $\sim T$, which is well
approximated even in the traditional approach,\cite{oliveira1981phaseshift,frota1986,sakai1989,costi1991,costi1994}
there is little difference in the results obtained either way. We
have explicitly calculated the spin thermopower for a set of parameters
$B$ and $T$ using the FDM NRG and compared them against the results
based on Eq.~(\ref{eq:transintnrg}); the difference is smaller than
the linewidth in the plot (a few percent at most), see Figs.~\ref{S_BT1}(b)
and \ref{S_BT1}(d). The FDM NRG is significantly slower than our
approach because it requires one calculation for each value of $T$
and, furthermore, because each such calculation is significantly slower
than a single calculation based on Eq.~(\ref{eq:transintnrg}). The
trade-off between an additional error of a few percent and an improvement
in the calculation efficiency of more than two orders of the magnitude
is well justified.

\section{\label{sec:Results}Results}

\begin{figure}
\begin{centering}
\includegraphics[width=1\columnwidth]{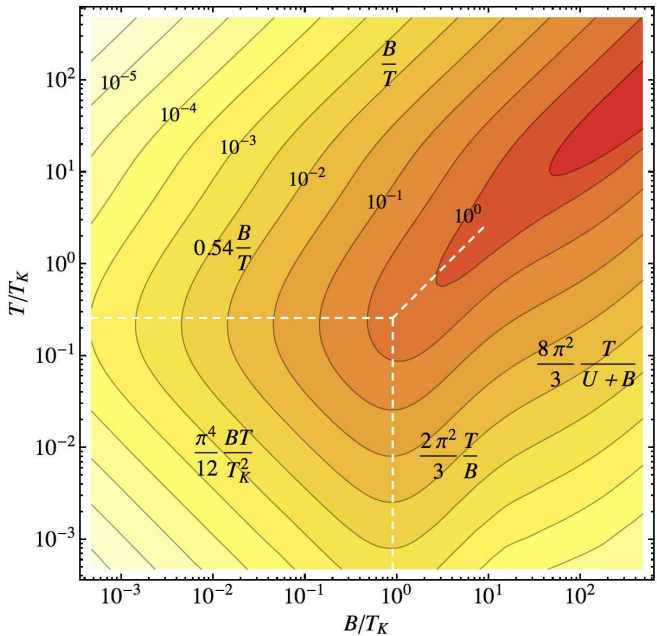} 
\par\end{centering}

\caption{\label{S_BT}Temperature and magnetic field dependence of the spin-Seebeck
coefficient (in units of $k_{B}/\left|e\right|$) as calculated with
the NRG method. The positions of the maxima at $B=\frac{2\sqrt{2}}{\pi}T_{K}$,
$T=0.26T_{K}$, and $T=0.29B$ are indicated with dashed lines. }
\end{figure}
We now turn to the numerical results for an interacting quantum dot.
We choose a strongly correlated regime with $U=8\Gamma$. The Kondo
temperature, Eq.~(\ref{eq:Tk}), at the particle-hole symmetric point
is then $T_{K}=0.11\Gamma$. 

In Fig.~\ref{S_BT} the spin-Seebeck coefficient is plotted for a
range of temperatures and magnetic fields covering all the regimes
of the symmetric Anderson model. It is evident from the Figure that
for temperatures and magnetic fields at or above the Kondo temperature,
but of roughly the same size, there is a ridge of high spin-Seebeck
coefficient (of the order of $k_{B}/\left|e\right|$). At the point
where the temperature and the field are both of the order of the Kondo
temperature this ridge splits into two: one at the temperature of
the order of the Kondo temperature with the spin-Seebeck coefficient
decaying linearly as the field vanishes and the other at the magnetic
field of the order of the Kondo temperature, again decaying linearly
as the temperature vanishes. The ridges separate three slopes where
the spin-Seebeck coefficient vanishes in directions away from the
ridges.

\begin{figure}
\begin{centering}
\includegraphics[width=1\columnwidth]{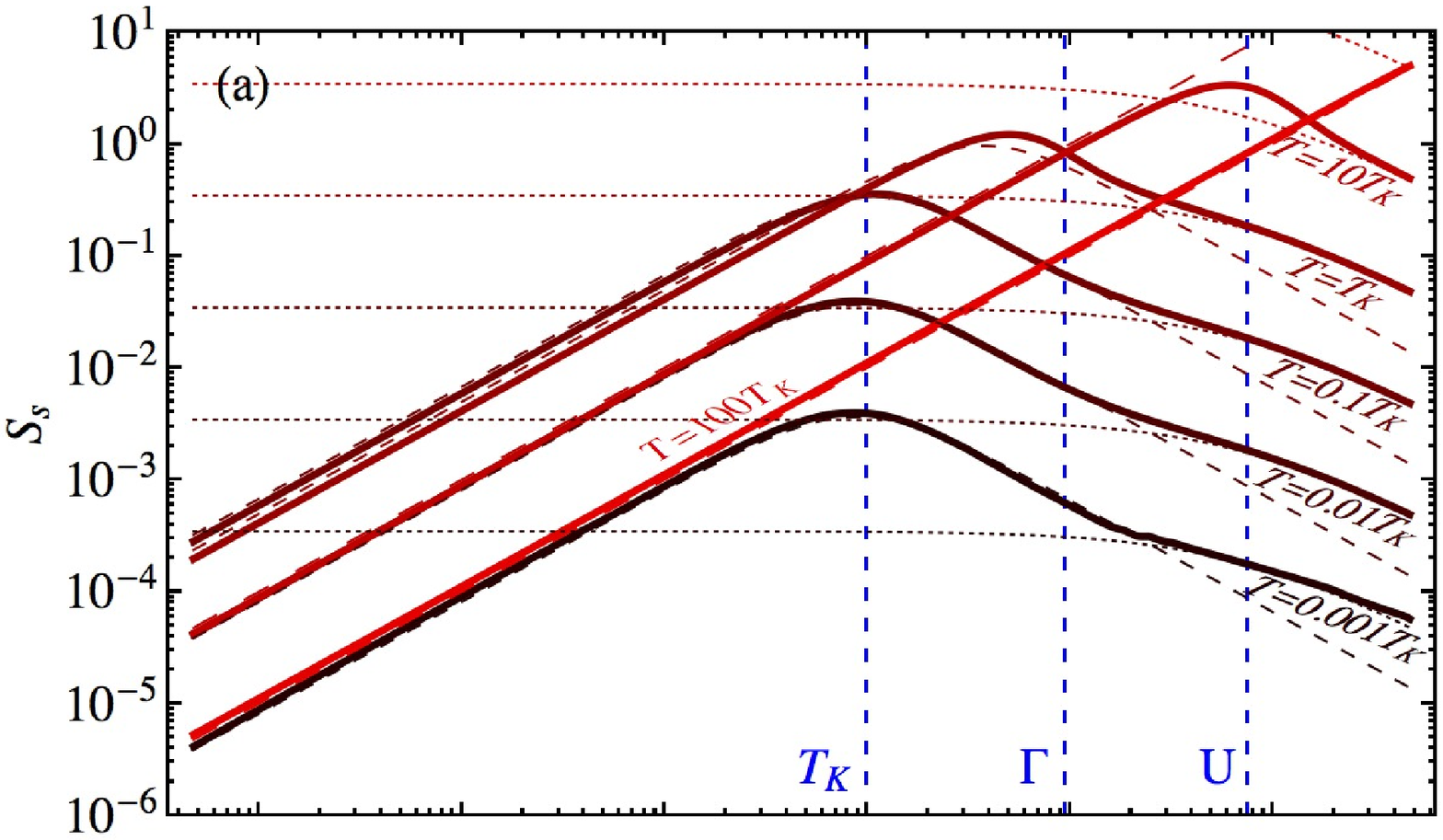}
\par\end{centering}

\begin{centering}
\includegraphics[width=1\columnwidth]{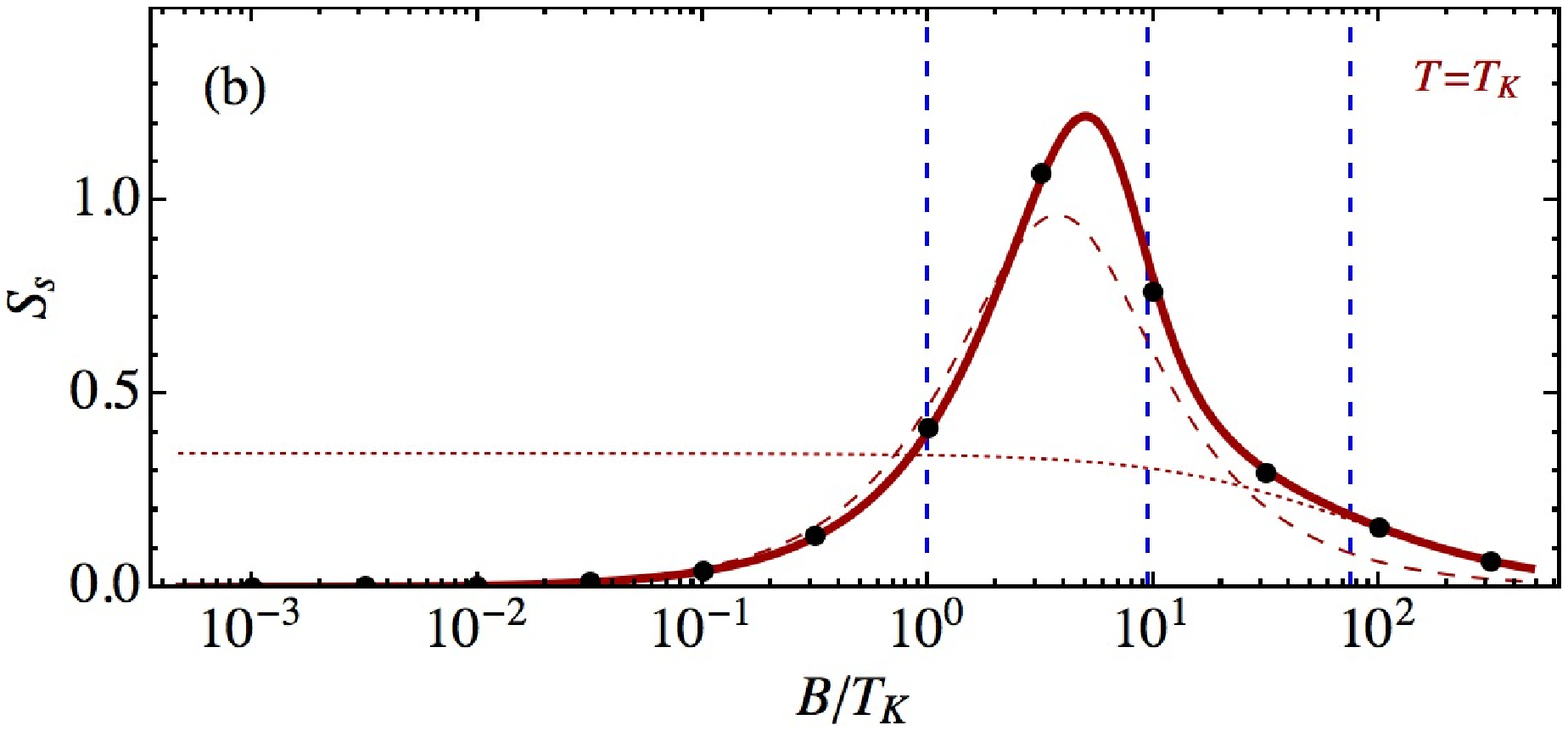} 
\par\end{centering}

\begin{centering}
\includegraphics[width=1\columnwidth]{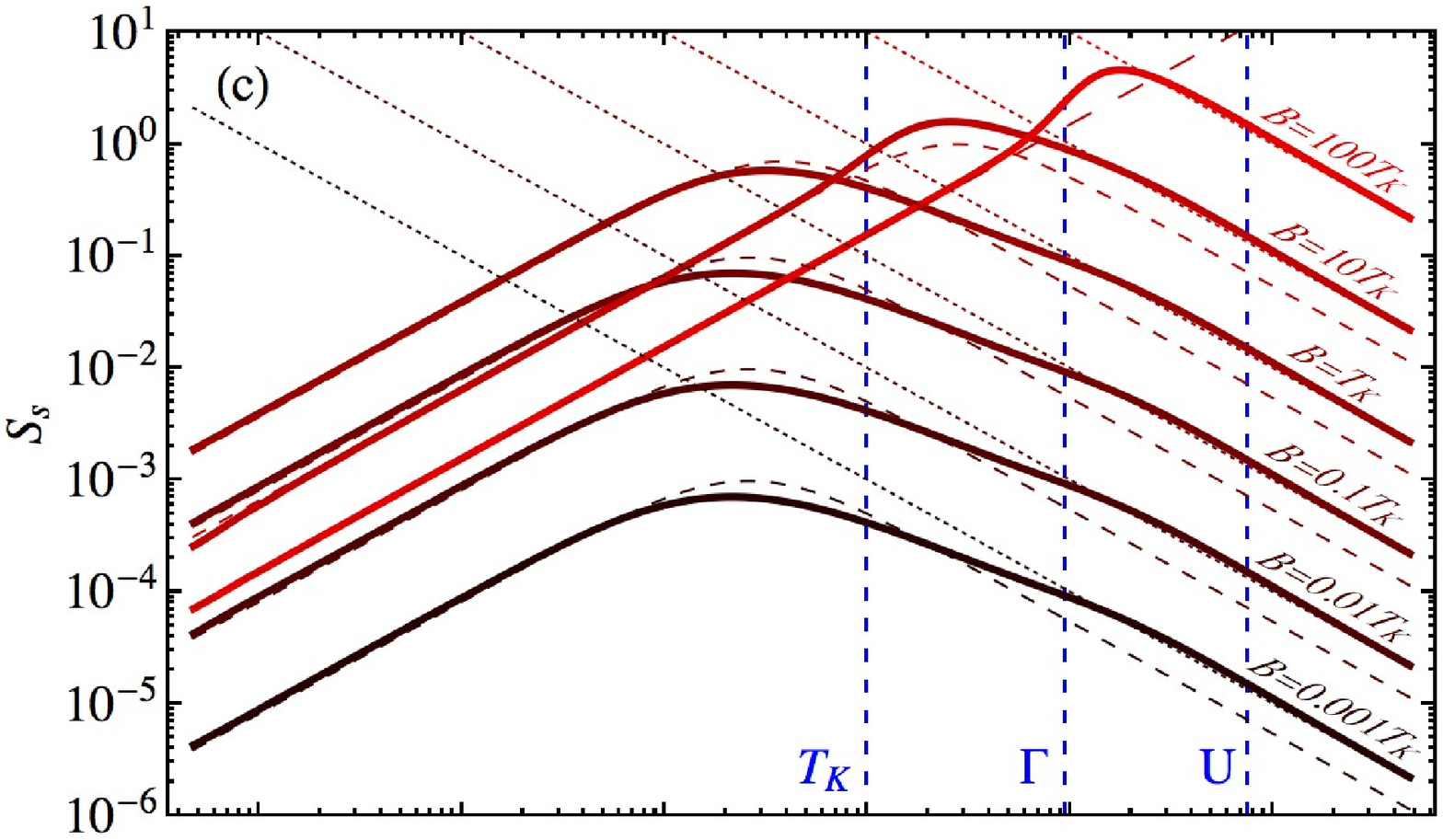}
\par\end{centering}

\begin{centering}
\includegraphics[width=1\columnwidth]{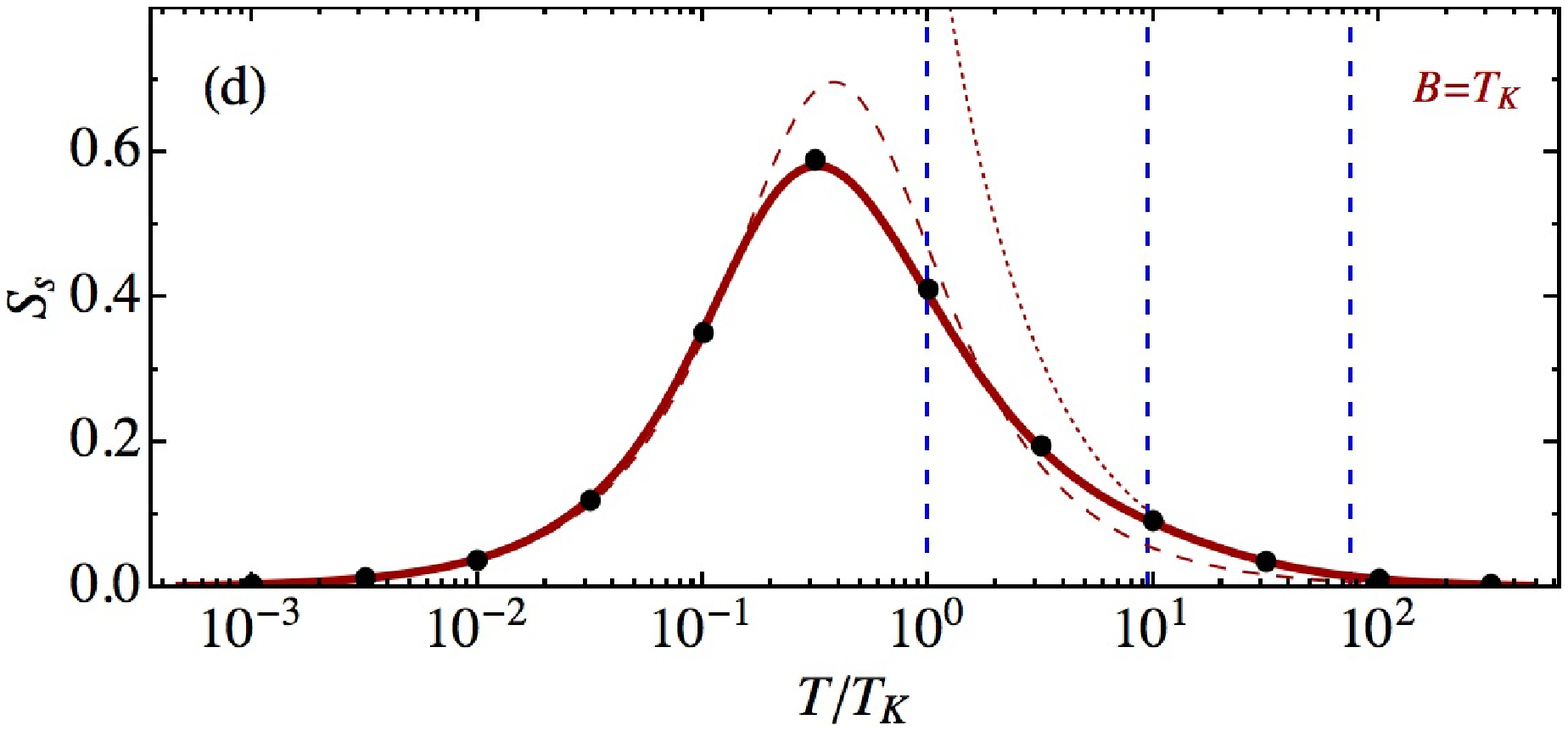} 
\par\end{centering}

\caption{\label{S_BT1}(a) Magnetic field dependence of the spin-Seebeck coefficient
for several values of the temperature. The low magnetic field approximations
are shown with dashed lines for $T=0.001T_{K}$, $T=0.01T_{K}$, $T=0.1T_{K}$
and $T=T_{K}$ {[}Eqs.~(\ref{eq:exact_NI}) and (\ref{eq:width_BT}){]}
and with long dashed lines for $T=10T_{K}$ and $T=100T_{K}$ {[}Eq.~(\ref{eq:seebeck_lowB}){]}.
The high magnetic field asymptotics, Eq.~(\ref{eq:Seebeck_lowT_highB}),
are shown with dotted lines. (c) Temperature dependence for several
values of the magnetic field. The low-temperature approximations are
shown with dashed lines for $B=0.001T_{K}$, $B=0.01T_{K}$, $B=0.1T_{K}$,
$B=T_{K}$ and \textbf{$B=10T_{K}$} {[}Eqs.~(\ref{eq:exact_NI})
and (\ref{eq:width_BT}){]} and with a long dashed line for $B=100T_{K}$
{[}Eq.~(\ref{eq:Seebeck_lowT_highB}){]}. The high-temperature asymptotics,
Eq.~(\ref{eq:seebeck_lowB}), are shown with dotted lines. In (b)
and (d) the spin-Seebeck coefficient is presented in linear scale
for $T=T_{K}$ and $B=T_{K}$, respectively. Black dots show the results
obtained using the FDM NRG method. }
\end{figure}
We show that a treatment similar to that for a noninteracting quantum
dot in Sec.~\ref{sec:Noninteracting-quantum-dot} is possible in
the Kondo regime (and its vicinity) provided that we account for the
temperature and field dependence of the position and the width of
the Kondo resonance. On top of that we find simple asymptotic expressions
for the spin thermopower far away from the Kondo regime. In Fig.~\ref{S_BT1}(a)
we compare these analytical results to the NRG data for various temperatures
as a function of magnetic field on a log-log plot. In Fig.~\ref{S_BT1}(b)
one set of data is reproduced on a log-linear plot. In Figs.~\ref{S_BT1}(c)
and (d) we present such comparisons for various values of the magnetic
field as a function of temperature. In the regions of their validity,
the analytical expressions are found to reproduce the exact values
reasonably well.

\subsection*{Particle-hole symmetric point, $\boldsymbol{\delta=0}$}

We first study low temperature $T\ll\max\left(T_{K,}B\right)$ Fermi-liquid
regimes and low magnetic field $B\ll\max\left(T_{K},T\right)$ regimes
of the particle-hole symmetric model separately. Then we combine the
results and present a unified theory which is capable of describing
the spin thermopower at $B,T\sim T_{K}$.

\subsubsection*{Low-temperature regimes, $T\ll\max\left(T_{K,}B\right)$}

Due to many-body effects there is an energy scale induced in the Anderson
model which plays the role of an effective bandwidth. This energy
scale is the Kondo temperature $T_{K}$ at low magnetic fields $B\ll T_{K}$
and rises up to $\Gamma$ as the magnetic field kills the Kondo effect.
At temperatures low compared to this energy scale, \textit{i.e.},
in the Fermi-liquid regime, only the spectral function in the vicinity
of the Fermi level is relevant to the calculation of the transport
properties. Here the Green's function can be parametrized in terms
of the Fermi-liquid quasiparticle parameters, $G_{\sigma}\left(\omega\right)=z_{\sigma}/(\omega-\tilde{\varepsilon}_{d\sigma}+i\tilde{\Gamma}_{\sigma})$,
where $z_{\sigma}=\left[1-\Sigma_{\sigma}^{\prime}\left(\mu\right)\right]^{-1}$
is the wavefunction renormalization factor, while $\tilde{\varepsilon}_{d\sigma}=z_{\sigma}\left[\varepsilon_{d}+\sigma\frac{B}{2}+\Sigma_{\sigma}\left(\mu\right)\right]$
and $\tilde{\Gamma}_{\sigma}=z_{\sigma}\Gamma$ are the quasiparticle
energy level and its half-width, respectively.\cite{Hewson93} This
parametrization provides an accurate description of the behavior of
the system in the vicinity of the Fermi level, thus it is suitable
for studying the transport properties; however, the features in the
spectral function away from the Fermi level are not well described.

The spin-Seebeck coefficient is derived by performing the Sommerfeld
expansion of the transport integrals in Eq.~(\ref{eq:S}), resulting
in the spin analog to the Mott formula, 
\begin{equation}
S_{s}=\frac{2\pi^{2}}{3}\frac{A_{\uparrow}^{\prime}\left(0\right)}{A_{\uparrow}\left(0\right)}T=\frac{4\pi^{2}}{3}\frac{\tilde{\varepsilon}_{d\uparrow}T}{\tilde{\varepsilon}_{d\uparrow}^{2}+\tilde{\Gamma}_{\uparrow}^{2}}.\label{eq:seebeck_lowT}
\end{equation}

In the strong coupling regime, $B\ll T_{K}$, the quasiparticle level
is shifted away from the chemical potential, $\tilde{\varepsilon}_{d\sigma}=\sigma\frac{B}{2}R$,
where the Wilson ratio is $R=2$ due to residual quasiparticle interaction\cite{Hewson06}
while its half-width is of the order of the Kondo temperature, $\tilde{\Gamma}_{\sigma}=\frac{4}{\pi}T_{K}$.\cite{Yamada75,Yamada75a}
Thus the spin-Seebeck coefficient in the $B\ll T_{K}$ and $T\ll T_{K}$
regime increases linearly with both the temperature and the magnetic
field: 
\begin{equation}
S_{s}=\frac{\pi^{4}}{12}\frac{BT}{T_{K}^{2}}.\label{eq:Seebeck_lowT_lowB}
\end{equation}

In the intermediate regime, $T_{K}\lesssim B\ll U$, the magnetic
moment is localized but, due to strong logarithmic corrections, the
field dependence of $\tilde{\epsilon}_{d\sigma}$ and $\tilde{\Gamma}_{\sigma}$
is nontrivial.\cite{Hewson06} Anyway, a good agreement with numerical
data is obtained for $B\lesssim\Gamma$ by allowing for the broadening
of the quasiparticle level due to the magnetic field,\cite{Costi00,Andrei82}
\begin{equation}
\tilde{\Gamma}_{\sigma}^{2}=\left(\frac{4}{\pi}T_{K}\right)^{2}+B^{2},\label{eq:width_B}
\end{equation}
 and keeping the zero-field expression for the level shift, $\tilde{\epsilon}_{d\sigma}=\sigma B$:
\begin{equation}
S_{s}=\frac{4\pi^{2}}{3}\frac{BT}{2B^{2}+\left(\frac{4}{\pi}T_{K}\right)^{2}}.\label{eq:Seebeck_maxB}
\end{equation}
 This expression reaches its maximal value at $B=\frac{2\sqrt{2}}{\pi}T_{K}$.
In the $T_{K}\ll B\ll\Gamma$ regime the spin-Seebeck coefficient
is inversely proportional to the magnetic field: 
\begin{equation}
S_{s}=\frac{2\pi^{2}}{3}\frac{T}{B}.\label{eq:Seebeck_lowT_midB}
\end{equation}

In the high magnetic field regime, $B\gtrsim U$, the Kondo resonance
is no longer present and the Hubbard I approximation to the Green's
function may be used,\cite{Hubbard64} 
\begin{equation}
G_{\text{\ensuremath{\sigma}}}\left(\omega\right)=\frac{1-\left\langle n_{d,-\sigma}\right\rangle }{\omega-\epsilon_{d}-\sigma\frac{B}{2}+i\Gamma}+\frac{\left\langle n_{d,-\sigma}\right\rangle }{\omega-\epsilon_{d}-U-\sigma\frac{B}{2}+i\Gamma},\label{eq:hubbard}
\end{equation}
 where $\left\langle n_{d\sigma}\right\rangle $ is the occupation
of the dot level with spin $\sigma$ electrons. In this regime, $\left\langle n_{d\downarrow}\right\rangle $
and $\left\langle n_{d\uparrow}\right\rangle $ approach 1 and 0,
respectively, and $\tilde{\varepsilon}_{d\sigma}\approx\sigma\left(\frac{U}{2}+\frac{B}{2}\right)$,
$\tilde{\Gamma}_{\sigma}=\Gamma$. The spin-Seebeck coefficient increases
linearly with temperature, but is inversely proportional to the magnetic
field for $B\gg U$: 
\begin{equation}
S_{s}=\frac{8\pi^{2}}{3}\frac{T}{U+B}.\label{eq:Seebeck_lowT_highB}
\end{equation}
In Fig.~\ref{S_BT1} we compare this expression to the NRG data and
find an excellent agreement in the appropriate regime.

\subsubsection*{Low magnetic field regimes, $B\ll\max\left(T_{K},T\right)$}

At low and intermediate temperatures, $T\ll\Gamma$, only the Kondo
resonance will play a role in determining the value of the spin-Seebeck
coefficient. The effective dot level width, being equal to $\frac{4}{\pi}T_{K}$
at low temperatures, $T\ll T_{K}$, increases with temperature,\cite{Yamada75,Yamada75a}
\begin{equation}
\tilde{\Gamma}_{\sigma}^{2}=\left(\frac{4}{\pi}T_{K}\right)^{2}+\left(\pi T\right)^{2}.\label{eq:width_T}
\end{equation}
 Putting this expression, together with the low temperature expression
for the level shift $\tilde{B}=B$, into the noninteracting quantum
dot formula Eq.~(\ref{eq:nonint_lowBx}), we recover the Fermi-liquid
asymptotic result Eq.~(\ref{eq:Seebeck_lowT_lowB}) for $T\ll T_{K}$,
while for $T\gg T_{K}$, as the width of the dot level becomes proportional
to temperature, an asymptotic formula different from that of a noninteracting
level, Eq.~(\ref{eq:nonint_lowB}), results: 
\begin{equation}
S_{s}=0.54\frac{B}{T}.\label{eq:Seebeck_midT}
\end{equation}
 Between these asymptotic regimes the spin-Seebeck coefficient reaches
its maximal value at $T=0.26T_{K}$.

In the free orbital regime, $T\gtrsim U$, the Hubbard I approximation,
Eq.~(\ref{eq:hubbard}), can be used again. Now, due to high temperature
$\left\langle n_{d\downarrow}\right\rangle =\left\langle n_{d\uparrow}\right\rangle \approx\frac{1}{2}$
and the effect of the magnetic field is only to shift the spectral
functions, 
\[
\left.A_{\sigma}\left(\omega\right)\right|_{B}\to\left.A_{\sigma}\left(\omega-\sigma\frac{B}{2}\right)\right|_{B=0}.
\]
 As $-f^{\prime}\left(\omega\right)\approx\frac{1}{4T}$ in the region
where the spectral density is appreciable, we get a simple expression
\[
S_{s}=\frac{2}{T}\frac{\int\mathrm{d}\omega\omega\left.A_{\uparrow}\left(\omega-\frac{B}{2}\right)\right|_{B=0}}{\int\mathrm{d}\omega\left.A_{\uparrow}\left(\omega-\frac{B}{2}\right)\right|_{B=0}}.
\]
 Taking into account that in the absence of the magnetic field the
spectral function is even, one can readily show that the spin-Seebeck
coefficient is given by 
\begin{equation}
S_{s}=\frac{B}{T},\label{eq:seebeck_lowB}
\end{equation}
 thus recovering the non-interacting expression Eq.~\ref{eq:nonint_lowB}.
In Fig.~\ref{S_BT1} we show that this expression correctly describes
the asymptotic behavior of the spin-Seebeck coefficient in this regime.

\subsubsection*{Unified theory for $B,T\ll\Gamma$ regimes}

Here the physics is governed by the Kondo resonance in the spectral
function and its remnants at temperatures and magnetic fields above
the Kondo temperature. A unified description of low and intermediate
temperature and magnetic field regimes, $B,T\ll\Gamma$, is obtained
using the noninteracting quantum dot spin-Seebeck coefficient formula,
Eq.~(\ref{eq:exact_NI}), and taking into account that the width
of the Kondo resonance depends on both the temperature and the field.
The appropriate phenomenological expression is a generalization of
the low temperature, Eq.~(\ref{eq:width_T}), and low magnetic field,
Eq.~(\ref{eq:width_B}), widths: 
\begin{equation}
\tilde{\Gamma}_{\sigma}^{2}=\left(\frac{4}{\pi}T_{K}\right)^{2}+B^{2}+\left(\pi T\right)^{2}.\label{eq:width_BT}
\end{equation}
Again we use $\tilde{B}=B$. As evident in Fig.~\ref{S_BT1}, this
gives quite an accurate approximation, reproducing the correct position,
width and, to a lesser extent, height of the peak in the spin-Seebeck
coefficient. It becomes even more accurate for $U=16\Gamma$ (not
shown here) where the Kondo energy scale is better separated from
higher energy scales, resulting in a better agreement with the NRG
data in the $B\lesssim T_{K}$ and $T_{K}\ll T\ll\Gamma$ regime.

At the point where the $B=\frac{2\sqrt{2}}{\pi}T_{K}$ and $T=0.26T_{K}$
maxima merge, the spin-Seebeck coefficient reaches a value of $S_{s}=0.53$,
while the current approximation gives $S_{s}=0.63$. This universal
value (provided the Kondo energy scale is well separated from higher
energy scales, $T_{K}\ll\Gamma$) is somewhat lower than that of a
noninteracting quantum dot which is a direct consequence of the widening
of the Kondo energy level. From the merging point the spin-Seebeck
coefficient increases only slightly, $S_{s}\sim1$, along the $T=0.29B$
line.

As in the noninteracting case, Eq.~(\ref{eq:interp_NI}), we can
provide an analytical approximation for the spin-Seebeck coefficient
by interpolating between the three asymptotic expressions of Eqs.~(\ref{eq:Seebeck_lowT_lowB}),
(\ref{eq:Seebeck_lowT_midB}) and (\ref{eq:Seebeck_midT}): 
\begin{equation}
S_{s}^{-1}=\left(\frac{\pi^{4}}{12}\frac{BT}{T_{K}^{2}}\right)^{-1}+\left(\frac{2\pi^{2}}{3}\frac{T}{B}\right)^{-1}+\left(0.54\frac{B}{T}\right)^{-1}.\label{eq:interp_I}
\end{equation}

\subsection*{Away from the particle-hole symmetric point, $\boldsymbol{\delta\ne0}$}

\begin{figure}
\begin{centering}
\includegraphics[width=1\columnwidth]{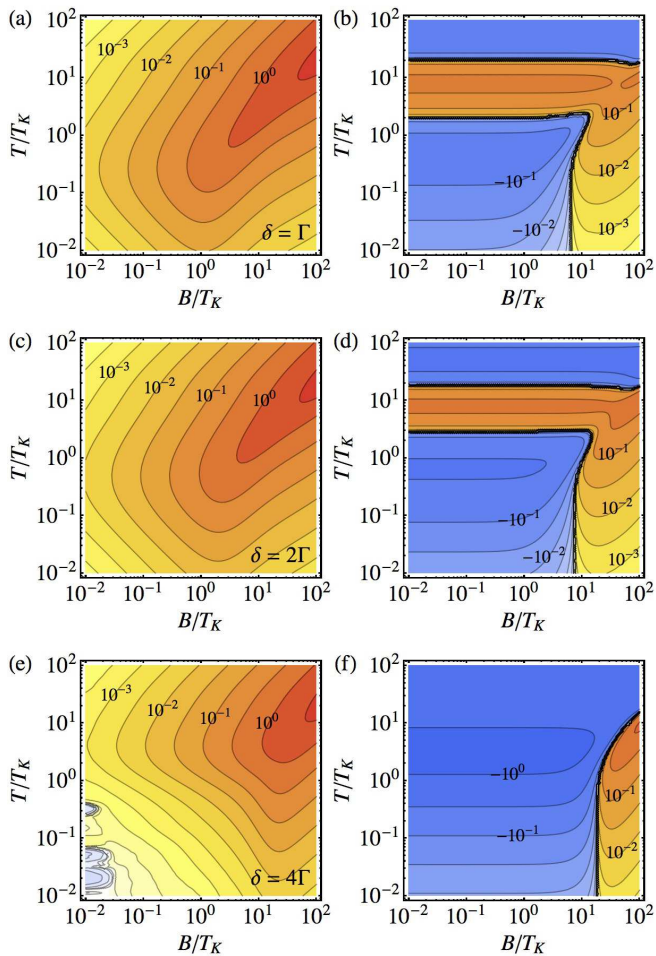} 
\par\end{centering}

\caption{\label{S_BT_away}Temperature and magnetic field dependence of the
spin-Seebeck coefficient (in units of $k_{B}/\left|e\right|$) as
calculated with the NRG method away from the particle-hole symmetric
point for (a) $\delta=\Gamma$, (c) $\delta=2\Gamma$ and (e) $\delta=4\Gamma$.
The corresponding electrical voltages, $-eV/\Delta T$, required for
electrical current to vanish are shown in (b), (d) and (f). Temperatures
and magnetic fields are shown in units of the Kondo temperature in
the particle-hole symmetric point. Note that because of the increase
of the Kondo temperature with $\delta$, the maxima are shifted to
higher temperature and magnetic field values. The low temperature
and magnetic field structure in (e) is an artifact of the numerical
method.}
\end{figure}
In Fig.~\ref{S_BT_away} we present the spin-Seebeck coefficient
for three different values of the asymmetry parameter $\delta$ ranging
from the Kondo to the mixed valence regime. In the Kondo regime, $\delta=\Gamma=\frac{U}{8}$
in Fig.~\ref{S_BT_away}(a) and $\delta=2\Gamma=\frac{U}{4}$ in
Fig.~\ref{S_BT_away}(b), the Kondo peak in the spectral function
is pinned to the chemical potential in the absence of the magnetic
field, $\tilde{\epsilon}_{d\sigma}\ll\tilde{\Gamma}_{\sigma}$.\cite{Hewson04}
The Kondo temperature increases with $\delta$, $T_{K}\propto\exp\left(\pi\delta^{2}/2U\Gamma\right)$,\cite{Hewson93}
which causes the positions of maxima in the spin-Seebeck coefficient
to shift to higher temperatures and fields. Provided we take this
effect into account, the behavior of the spin-Seebeck coefficient
reproduces that of the particle-hole symmetric situation in Fig.~\ref{S_BT}.
In the mixed valence regime, $\delta=4\Gamma=\frac{U}{2}$ in Fig.~\ref{S_BT_away}(e),
the Kondo peak has merged with the atomic peak resulting in a resonance
in the spectral function of width $\tilde{\Gamma}_{\sigma}\sim\Gamma$
at $\tilde{\epsilon}_{d\sigma}\sim\tilde{\Gamma}_{\sigma}$ at $B=0$.\cite{Haldane78}
The maxima shift to $T,B\sim\Gamma$ and we reproduce the characteristic
suppression of the spin-Seebeck coefficient at low temperatures and
fields observed in the noninteracting case in Fig.~\ref{S0_BT_away}(c).
In the empty orbital regime (not shown) the peak of width $\Gamma$
would shift even further away from the chemical potential and we would
reproduce the noninteracting result of Fig.~\ref{S0_BT_away}(e).

The electrical voltage required to stop the electrical current matches
the prediction of the noninteracting theory in the mixed valence regime,
Fig.~\ref{S_BT_away}(f), while in the Kondo regime, Figs.~\ref{S_BT_away}(b)
and (d), it changes sign twice as a function of temperature at low
magnetic fields. Such a behavior is consistent with the temperature
dependence of the charge thermopower of a quantum dot studied by Costi
and Zlatić in Ref.~{[}\onlinecite{Costi10a}{]}.

\section{\label{sec:Summary}Summary}

We demonstrated that the spin-Seebeck effect in a system composed
of a quantum dot in a magnetic field, attached to paramagnetic leads,
can be utilized to provide a pure spin current for spintronic applications
provided the dot is held at the particle-hole symmetric point. By
tuning the temperature and the field, the spin-Seebeck coefficient
can reach large values of the order of $k_{B}/\left|e\right|$. Namely,
for temperatures higher than $0.26T_{K}$, such a large spin thermopower
is available at the magnetic field where $T=0.29B$. We carefully
analyzed different regimes and derived analytical formulae which successfully
reproduce and explain the temperature and magnetic field dependence
of the spin-Seebeck coefficient calculated numerically with NRG. 

Replacing the magnetic field with the gate voltage, the same conclusions
are also valid for the charge thermopower of a negative-$U$ quantum
dot, for which the charge-Seebeck coefficient of the order of $k_{B}/\left|e\right|$
was recently reported in Ref.~{[}\onlinecite{Andergassen11}{]}.

We also studied the spin-Seebeck coefficient away from the particle-hole
symmetric point. Our analysis shows that in the Kondo regime the results
do not change qualitatively provided we take into account the increase
of the Kondo temperature and apply a suitable electrical voltage across
the dot to stop the electrical current. 

As the Kondo temperature in quantum dots can be made quite low, the
magnetic fields required to reach the large spin thermopower regime
should be easily achievable in experiment. 
\begin{acknowledgments}
We acknowledge the support from the Slovenian Research Agency under
Contract No. P1-0044. 
\end{acknowledgments}

\section*{Appendix A}

\begin{figure}
\begin{centering}
\includegraphics[width=0.85\columnwidth]{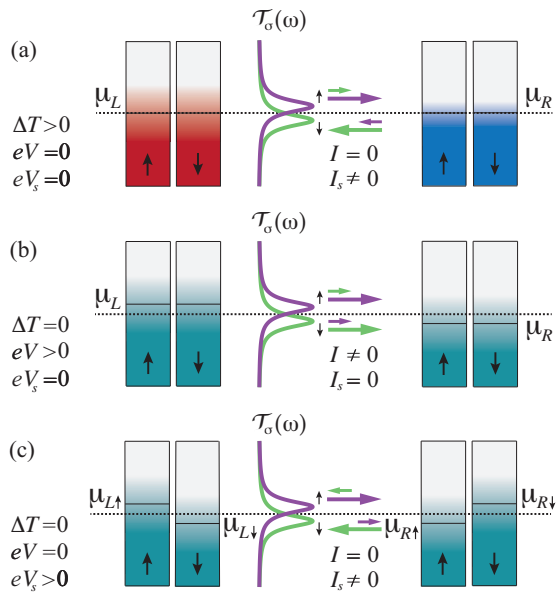} 
\par\end{centering}

\caption{\label{S_FigA}Currents generated by (a) temperature difference, (b)
electrical voltage and (c) spin voltage applied across a quantum dot
in a magnetic field $B>0$. The currents are shown separately for
spin-up (violet arrows) and spin-down electrons (green arrows), as
well as for electrons with energies higher (upper set of arrows) and
lower (lower set of arrows) than the equilibrium chemical potential
(black dotted line).}
\end{figure}
According to Eq.~(\ref{eq:I_sigma}) the electrical current of spin
$\sigma$ electrons is determined by the energy distribution of incoming
electrons $f_{L\sigma}\left(\omega\right)-f_{R\sigma}\left(\omega\right)$
and the probability that those electrons are transmitted through the
quantum dot ${\cal T}_{\sigma}\left(\omega\right)$. In the presence
of a magnetic field, ${\cal T}_{\sigma}\left(\omega\right)$ is asymmetric
about the chemical potential. Assuming the particle-hole symmetry
and $B>0$ it is larger for spin-up electrons immediately above the
chemical potential ($\omega>0$) than for those immediately below
the chemical potential ($\omega<0$), and vice versa for spin-down
electrons as ${\cal T}_{\downarrow}\left(\omega\right)={\cal T}_{\uparrow}\left(-\omega\right)$. 

In the linear response regime we can study the effects of $\Delta T$,
$eV$ and $eV_{s}$ separately as follows. 

In the presence of a temperature difference $\Delta T$, Fig.~\ref{S_FigA}(a),
the energy distribution of the incoming electrons is the same for
spin-up and spin-down electrons. For $\Delta T>0$ the incoming electrons
with $\omega>0$$ $ originate from the left (hot) lead while those
with $\omega<0$ originate from the right (cold) lead. Now for $\omega>0$
there will be a surplus of spin-up electrons reaching the right lead
due to the asymmetry of ${\cal T}_{\sigma}\left(\omega\right)$ while
for $\omega<0$ the same surplus of spin-down electrons will reach
the left lead, resulting in a zero electrical current and a finite
spin current across the dot. 

In the presence of an electrical voltage $eV>0$, Fig.~\ref{S_FigA}(b),
all the incoming electrons originate from the left lead, their energy
distribution is an even function of $\omega$ and is again the same
for both spin projections. For $\omega>0$ a surplus of spin-up electrons
is reaching the right lead while for $\omega<0$ the same surplus
of spin-down electrons is reaching the right lead. The result is a
finite electrical current and a zero spin current. 

In the presence of a spin voltage $eV_{s}>0$, Fig.~\ref{S_FigA}(c),
spin-up electrons originate from the left lead and spin-down electrons
originate from the right lead. Energy distributions of the two species
of incoming electrons, like the transmission probabilities ${\cal T}_{\sigma}\left(\omega\right)$,
are related by reflection symmetry with respect to $\omega=0$. Consequently,
the number of spin-up electrons reaching the right lead is the same
as the number of spin-down electrons reaching the left lead, \textit{i.e.}
a zero electrical current and a finite spin current.

\bibliography{papers}

\end{document}